\documentclass[conference]{IEEEtran}
\IEEEoverridecommandlockouts
\usepackage{cite}
\usepackage{amsmath,amssymb,amsfonts,bm}
\usepackage{algorithm}
\usepackage[noend]{algpseudocode}
\usepackage{nopageno}
\usepackage{graphicx}
\usepackage{textcomp}
\usepackage{hyperref}
\usepackage{xcolor}
\usepackage[caption=false]{subfig}
\usepackage{comment}
\definecolor{niceblue}{rgb}{0, 0.5, 1.0}

\hypersetup{
    colorlinks=true,
    linkcolor=niceblue,
    filecolor=magenta,      
    urlcolor=niceblue,
    citecolor=niceblue,
    pdftitle={Overleaf Example},
    pdfpagemode=FullScreen,
    }

\begin{document}

\title{Optimization-Based Exploration of the Feasible Power Flow Space for Rapid Data Collection\\
\thanks{This work is supported in part by the HORIZON-MSCA-2021 Postdoctoral Fellowship Program, Project \#101066991 -- TRUST-ML.}
}

\author{\IEEEauthorblockN{Ignasi Ventura Nadal and Samuel Chevalier}
\IEEEauthorblockA{Department of Wind and Energy Systems \\
\textit{Technical University of Denmark}\\
Kongens Lyngby, Denmark \\
\{s212393,schev\}@dtu.dk}
}

\IEEEoverridecommandlockouts
\IEEEpubid{\makebox[\columnwidth]{978-1-6654-3254-2/22/\$31.00~\copyright2022 IEEE \hfill}\hspace{\columnsep}\makebox[\columnwidth]{ }}

\maketitle

\IEEEpubidadjcol

\pagestyle{plain}
\section*{Abstract}
\begin{abstract}
This paper provides a systematic investigation into the various nonlinear objective functions which can be used to explore the feasible space associated with the optimal power flow problem. A total of 40 nonlinear objective functions are tested, and their results are compared to the data generated by a novel exhaustive rejection sampling routine. The Hausdorff distance, which is a min-max set dissimilarity metric, is then used to assess how well each nonlinear objective function performed (i.e., how well the tested objective functions were able to explore the non-convex power flow space). Exhaustive test results were collected from five PGLib test-cases and systematically analyzed.
\end{abstract}


\section{Introduction}
Recently, many challenging problems associated with the operation and control of electric power systems have benefited from learning-based and data-driven approaches. Machine learning (ML) in particular is helping to transform a number of otherwise intractable problems into ones which can now be tractably solved; such examples include solving unit commitment with surrogate frequency nadir constraints~\cite{Zhang:2022}, performing chance-constrained outage scheduling considering real-time market balancing~\cite{Dalal:2019}, and encoding small-signal stability constraints into the optimal power flow (OPF) problem~\cite{Murzakhanov:2020}. Many other learning-based success stories are reviewed in recent literature surveys which focus on energy system reliability management~\cite{Duchesne2020} and reinforcement learning applications~\cite{Chen:2021}.

Despite the vastly different applications and algorithmic technologies associated with each of these research works, they are all united by a singular commonality: the desperate need for high-quality and properly sampled \textit{training data}. Accordingly, a number of recent works in the field of power systems have focused on database generation. For instance, advanced importance sampling is employed in~\cite{Krishnan:2011} for generating probabilistic security assessment data; a two-stage procedure biases the sampling routine towards data in ``high information content" regions. Similarly, ``directed walk" methods, which iteratively push operating points towards small-signal stability boundaries, have been employed for power system security assessment~\cite{Thams:2020} and offshore energy island controller tuning~\cite{Stiasny:2022}.

To accelerate sampling, many works have utilized relaxation-based methods for quickly classifying large regions of the power flow space as infeasible. To perform classification,~\cite{Molzahn:2017} developed an SOCP and Lasserre hierarchy-based hypersphere ``grid pruning'' algorithm. This pruning procedure was extended in~\cite{Thams:2020}, where SDP-based hyperspheres classified regions of infeasibility, thus enabling rapid rejection sampling. Quadratic convex OPF relaxations were utilized in~\cite{Venzke:2021} to generate certificates of infeasibility via separating hyperplanes. A Monte Carlo sampling approach, known as hit and run sampling, was utilized by the OPF-Learn tool~\cite{Jones:2021}, thus allowing for rapid sampling of high dimensional polytopes formed by SOCP relaxation. Other optimization-based techniques have focused on directly characterizing the feasible OPF space. Authors in~\cite{Xue:2020} and~\cite{Chiang:2018} used trajectory unified and quotient gradient system methods, respectively, for feasible space characterization of small power systems. In order to discover the drivers of convexity gaps, the nonconvexities associated with various OPF problems were empirically investigated in~\cite{Narimani:2018}.

All of the OPF-related methods surveyed by the authors focus on either (i) classifying regions of the OPF space as feasible or not, thus accelerating rejection sampling, or (ii) directly characterizing the feasible space with a surrogate model. To the author's knowledge, no published methods have directly investigated how a nonlinear objective function can be used to optimally explore the feasible space. In this paper, we do just this: we use an optimization solver (IPOPT) to maximize an objective function which is rewarded for finding OPF solutions which are maximally far apart. Specifically, our goal is to systematically study the many different objective functions which can be used for finding maximally spread-out OPF solutions. Following are our primary contributions:
\begin{enumerate}
\item We define an optimization-based data collection routine (Alg.~\ref{algo:iterative}), and then we test 40 competing nonlinear objective functions which explore the feasible space.
\item We define a separate exhaustive rejection sampling routine (Alg.~\ref{algo:exhaustive}) to discretely approximate the feasible space, and we show how the Hausdorff distance can use the results of Algs. \ref{algo:iterative} and \ref{algo:exhaustive} to assess how well the various objective functions are able to cover the feasible space.
\item We post our code~\cite{github}, so that others in the power flow community can either (i) test their own objective functions for feasible space exploration or (ii) directly use the ones we have found to be most successful.
\end{enumerate}
The remainder of this paper is structured as follows. In Sec. \ref{sec:II}, we state a common OPF model, and we define various nonlinear objective functions which can explore the feasible space. In Sec. \ref{sec:III}, we build an exhaustive sampling algorithm which approximates the feasible space, and we show how the Hausdorff distance can be used to assess feasible space coverage by the nonlinear objective functions. Test results, collected from five simulated test systems are presented in Sec. \ref{sec:IV}, and Conclusions are offered in Sec. \ref{sec:IV}

\section{Opf data collection via space exploration}\label{sec:II}
In this section, we first state a common OPF formulation, and then we introduce an associated objective function which, over many iterations, will optimally explore the feasible space. Finally, we introduce the family of functions which we test in this paper for optimally exploring the feasible space.

\subsection{OPF formulation}
The OPF problem aims to determine the best-operating settings of a power system subject to the equality and inequality constraints which encode the system's physical limitations and operational characteristics. To state the OPF problem, we consider a network with a set of buses ${\mathcal N} \triangleq \{1,2,...,n\}$, a subset of which are dispatchable generators ${\mathcal G}\subset {\mathcal N}$, and a set of lines ${\mathcal L}\subset {\mathcal N}\times {\mathcal N}$. In this network, nodal voltage phasors are denoted by ${\bm v}e^{j{\bm \theta}}\in {\mathbb C}^{n\times1}$, and complex nodal power injections are given by ${\bm p}+j{\bm q}\in {\mathbb C}^{n\times1}$. These are related through the nonlinear and non-convex polar power flow equations:
\begin{subequations}\label{eq: inj}
\begin{align}\label{p_inj}
p_{i} & =v_{i}\sum_{k\in\mathcal{K}_{i}}v_{k}\left(G_{ik}\cos(\theta_{ik})+B_{ik}\sin(\theta_{ik})\right)\\
q_{i} & =v_{i}\sum_{k\in\mathcal{K}_{i}}v_{k}\left(G_{ik}\sin(\theta_{ik})-B_{ik}\cos(\theta_{ik})\right),\label{q_inj}
\end{align}
\end{subequations}
where ${\mathcal K}_i$ is the set of buses attached to bus $i$. 
Using (\ref{eq: inj}), the following OPF constraint model~\cite{lavaei_zero_2012} is developed in a per-unit system and is consistent with the PowerModels framework~\cite{Coffrin:2018}:
\begin{subequations}\label{eq: OPF_constraints}
\begin{align}
v_{i}^{{\rm min}} & \leq v_{i}\leq v_{i}^{{\rm max}}, &  & \forall i\in\mathcal{N}\\
p_{i} & =p_{G,i}-p_{D,i}, &  & \forall i\in\mathcal{N}\\
q_{i} & =q_{G,i}-q_{D,i} &  & \forall i\in\mathcal{N}\\
p_{G,j}^{{\rm min}} & \leq p_{G,j}\leq p_{G,j}^{{\rm max}}, &  & \forall j\in\mathcal{G}\\
q_{G,j}^{{\rm min}} & \leq q_{G,j}\leq q_{G,j}^{{\rm max}}, &  & \forall j\in\mathcal{G}\\
\left|\theta_{i}-\theta_{j}\right| & \le\theta_{ij}^{{\rm max}}, &  & \forall\{i,j\}\in\mathcal{L}\\
|p_{ij}+jq_{ij}| & \le s_{ij}^{{\rm max}}, &  & \forall\{i,j\}\in\mathcal{L}\\
|p_{ji}+jq_{ji}| & \le s_{ji}^{{\rm max}}, &  & \forall\{i,j\}\in\mathcal{L}\\
\theta_{i} & =0, &  & i=1,
\end{align}
\end{subequations}
where $p_{i}, q_{i}$ are given in (\ref{eq: inj}), and $p_{ij}$, $q_{ij}$ are the active and reactive line flow equations given by (\ref{eq: inj}) when $\mathcal{K}_{i}=\{i,j\}$ (and the admittance entries are updated accordingly). Typically, the OPF objective function associated with constraint set (\ref{eq: OPF_constraints}) quantifies the operational cost of power generation, and an optimization solver is used to feasibly minimize this cost~\cite{low_convex_2014}. However, we have deliberately omitted this cost function, since we intend to replace it with an alternative function which sequentially explores the feasible space.

\subsection{Nonlinear objectives for space exploration}
The objective of the data collection algorithm is to swiftly obtain OPF solutions that can accurately represent the feasible space of the problem. To achieve this, we try to solve the following intractable optimization problem:
\begin{align}\label{eq: OPF_iterative}
\max\; & \{\text{distance between $N$ feasible solutions}\}\\
{\rm s.t.}\; &\eqref{eq: inj}-\eqref{eq: OPF_constraints}.\nonumber
\end{align}
In order to practically solve (\ref{eq: OPF_iterative}), the subsequent process is applied. An initial power flow solution is computed by setting the objective function in (\ref{eq: OPF_iterative}) to 0. Thus, the initial point does not depend on any objective function parameters. Once an initial solution is calculated, the algorithm starts calculating OPF solutions by finding a new solution which is the ``furthest" from all the previous solutions. This procedure is repeated a total of $N$ times, and it is summarized in Alg.~\ref{algo:iterative}.

\begin{algorithm}
\caption{Sequential Data Collection}\label{algo:iterative}

{\small \textbf{Require:}
OPF model, desired number of data points $N$

\begin{algorithmic}[1]

\State Find any feasible solution ${\bm s}^{\star}_1$ to \eqref{eq: inj}-\eqref{eq: OPF_constraints}

\State Set ${\mathcal S} = {\bm s}^{\star}_1$ and $i=2$

\For{$i\le N$ }

\State ${\bm s}_i^{\star}={\rm argmax} \;\{ {\rm dist}({\bm s}_i,{\mathcal S})\}$ s.t. \eqref{eq: inj}-\eqref{eq: OPF_constraints}

\State ${\mathcal S} \leftarrow [{\mathcal S}, \; {\bm s}_i^{\star}]$

\State $i = i+1$

\EndFor {\bf end}

\State \Return Library of OPF solutions $\mathcal S$

\end{algorithmic}}
\end{algorithm}

Line 4 of Alg.~\ref{algo:iterative} includes a distance function which quantifies the distance between a library of pre-computed numerical solutions $\bm S$ and the decision variable vector ${\bm s}_i$. The primary goal of this paper is to answer the following question: \textit{what sort of nonlinear objective function is most effective at maximizing the distance between points in the feasible OPF space?}

In this paper, we tested a total of 40 different objective functions. The complete list of tested functions is posted on our public GitHub \cite{github}, available at \url{https://github.com/ignasiven/Data-Collection-Algorithm}; a subset of these functions are listed in the Appendix\footnote{The functions enumerated in the Appendix are the top-performing ones.}. Each nonlinear function quantifies the difference between a decision variable vector and all previously computed variable vector solutions. The function list was developed through trial and error, and it is not necessarily definitive nor complete. Due to the very flexible implementation through PowerModels, functions can be suggested, added, or modified quite easily. For greater insight into how the data collection algorithm works, the original code is available at \cite{github}. We note that the data collection procedure can easily be extended to additionally treat loads as decision variables (e.g., to collect data associated with a unit commitment problem, where loads are time-varying).





\section{Assessing feasible space coverage}\label{sec:III}
In this section, we explain the methodology we use for assessing the ability of each tested objective function to approximate the feasible space. Initially, exhaustive rejection sampling approximates the feasible space. Next, a Hausdorff distance metric assesses feasible space coverage.


\subsection{Exhaustive Rejection Sampling}
To assess feasible space coverage, we begin by exhaustively sampling from the feasible space; this is demonstrated pictorially by the small orange dots in Fig. \ref{fig:Haus}. This procedure is initialized by taking the OPF feasible region and ``partitioning" it, also shown in Fig. \ref{fig:Haus}. Thereupon, numerous optimizations are run by looping over these partitions and searching for multiple feasible solutions within each. Every time a feasible solution is found, its numerical variable vector ${\bm x}^{\star}_i$ is added to the exhaustive dataset $\mathcal{X}_e$:
\begin{equation}\label{eq: Ex}
    \mathcal{X}_e \leftarrow [\mathcal{X}_e, \; {\bm x}^{\star}_i].
\end{equation}
We note that this exhaustive sampling approach is only ever intended to be applied to small test systems for the express purpose of studying how the various nonlinear objective functions behave. Successful nonlinear objective functions can then be used for data collection on larger systems.

\subsubsection{Bound Partitioning Algorithm}

The idea of partitioning the feasible space was inspired by common grid searches used in many fields, where a massive grid is partitioned into smaller subset search regions. By iteratively and exhaustively searching for solutions in each one of these spaces, we decrease the likelihood that we ``miss" solutions in any particular region.

\begin{figure}
\centering
\includegraphics[width=0.9\columnwidth]{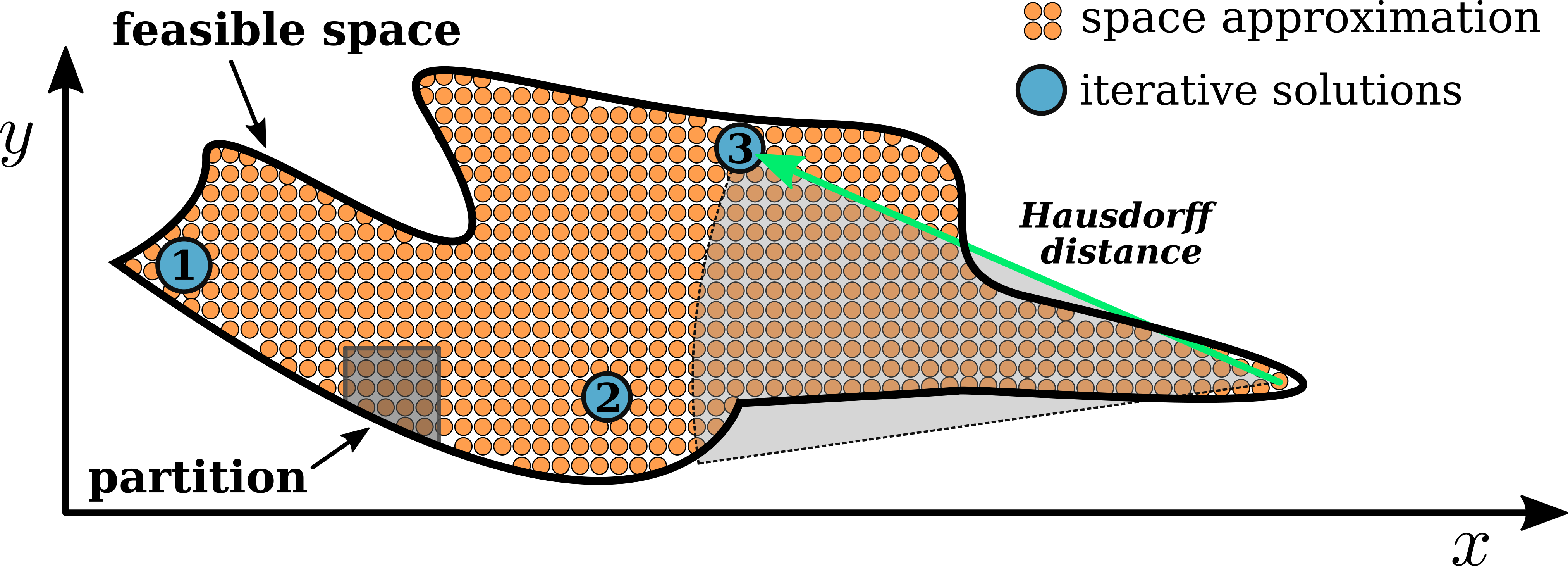}
\caption{Depicted is the Hausdorff distance (green arrow) associated with points collected in some 2-dimensional feasible space $f(x,y)$. The orange points exhaustively approximate the space, while the blue points are the sequential solutions of a given nonlinear objective function. The green arrow connects the orange point (farthest to the right) whose \textit{minimum distance} to an iterative solution is \textit{maximized}.}\label{fig:Haus}
\label{feasspace}
\end{figure}

Consider the following hypercube (or grid) associated with generator voltage limits:
\begin{equation}
    \mathcal{H}=[v_{G1}^{{\rm min}},v_{G1}^{{\rm max}}]\times\cdots\times[v_{Gn}^{{\rm min}},v_{Gn}^{{\rm max}}].
\end{equation}
After diving each voltage into $m$ equally sized ranges, we sequentially partition the hypercube by dividing it into $m^n$ different hypercubes with tighter voltage bounds. For example, if we partition each generator voltage limit into two smaller subsets (from ``min" to ``mid", and from ``mid" to ``max"), then the resulting $2^n$ partitions are given by
\begin{subequations}\label{eq: parts}
\begin{align}
\mathcal{H}_{1} & \!=[v_{G1}^{{\rm min}},v_{G1}^{{\rm {\bf mid}}}]\times[v_{G2}^{{\rm min}},v_{G2}^{{\rm {\bf mid}}}]\times[v_{G3}^{{\rm min}},v_{G3}^{{\rm {\bf mid}}}]\times\cdots\\
\mathcal{H}_{2} & \!=[v_{G1}^{{\rm {\bf mid}}},v_{G1}^{{\rm max}}]\times[v_{G2}^{{\rm min}},v_{G2}^{{\rm {\bf mid}}}]\times[v_{G3}^{{\rm min}},v_{G3}^{{\rm {\bf mid}}}]\times\cdots\\
\mathcal{H}_{3} & \!=[v_{G1}^{{\rm min}},v_{G1}^{{\rm {\bf mid}}}]\times[v_{G2}^{{\rm {\bf mid}}},v_{G2}^{{\rm max}}]\times[v_{G3}^{{\rm min}},v_{G3}^{{\rm {\bf mid}}}]\times\cdots\\
& \nonumber \quad\quad\quad\quad\quad\quad\quad\quad\quad\quad\vdots
\end{align}
\end{subequations}
Within each hypercube, there may exist many unique feasible OPF solutions; thus, it is crucial to consider all possible partitions to ensure no feasible solution is ignored. Once all voltage ranges have been defined, we loop over the generator voltage limit partitions ${\mathcal H}_i$ and attempt to find a feasible OPF within each by respecting the new voltage limits. An example partition is shown in Fig. \ref{fig:Haus}. 

Because the voltage ranges are highly restricted within each ${\mathcal H}_i$, the vast majority of partitions do not contain any feasible OPF solutions; through numerical study, we found that usually less than 2\% of the partitions contain a feasible solution. In order to increase the number of obtained solutions within each partition, whenever a feasible solution is found, the associated partition space is deemed feasible, and it is explored with an optimization function (just as described in the previous section). The choice of the associated objective function overlaps with the ultimate purpose of this paper. However, because it is not known a priori which sort of function will be the best to explore the feasible space, a function which prioritizes finding solutions which have different complex power generation values is used. We note that the chosen function in this stage is not as relevant as it is in Alg.~\ref{algo:iterative}, which attempts to explore the full power flow space in as few iterations as possible. In this case, (i) the feasible space is ``more bounded", so it is potentially easier to explore, and (ii) the routine can run for as many iterations as needed, since we are aiming for exhaustive space sampling. Hence, after many iterations, the dissimilarities between the alternative functions in the Appendix are not as significant as they are after several iterations (i.e., ``quantity overtakes quality").
Equation (\ref{fobje}) expresses the objective function used to exhaustively explore the partitions; it is equivalent to $f_3$ in the function list~\cite{github}:

\begin{footnotesize}
\begin{equation} \label{fobje}
    f({\mathcal X}_e)=\sum_{j\in\Gamma}\!\Bigg(\!\log\bigg(\sum_{i\in{\mathcal G}}\Big(p_{i}-P_{i,j}\Big)^{2}\bigg)+\log\bigg(\sum_{i\in{\mathcal G}}\Big(q_{i}-Q_{i,j}\Big)^{2}\bigg)\!\Bigg),
\end{equation}
\end{footnotesize}where $\Gamma$ is the set of all previously discovered data points in the exhaustive data set ${\mathcal X}_e$ from (\ref{eq: Ex}), $p_{i}$, $q_{i}$ are decision variables, and $P_{i,j}$, $Q_{i,j}$ are numerical values from ${\mathcal X}_e$. As (\ref{fobje}) is sequentially maximized (with respect to the OPF and partitioning constraint), the associated brute force solutions begin to create an accurate overall representation of the system (similar to the small orange dots in Fig. \ref{fig:Haus}). The full exhaustive rejection sampling routine is outlined in Alg.~\ref{algo:exhaustive}.

\begin{algorithm}
\caption{Exhaustive Rejection Sampling}\label{algo:exhaustive}

{\small \textbf{Require:}
OPF model, $M$ feasible space partitions ${\mathcal H}_i$, $T$ maximum desired OPF solutions within each partition

\begin{algorithmic}[1]

\State Set $i=1$

\For{$i\le M$ }

\If{\{\eqref{eq: inj}-\eqref{eq: OPF_constraints} $\cap$ ${\mathcal H}_i$\} are feasible,}

\State Set $j=1$

\For{$j\le T$ }

\State ${\bm x}^{\star}={\rm argmax} \;\{ \eqref{fobje} \}$ s.t. \{\eqref{eq: inj}-\eqref{eq: OPF_constraints} $\cap$ ${\mathcal H}_i$\}

\State ${\mathcal X}_e \leftarrow [{\mathcal X}_e, \; {\bm x}^{\star}]$

\State $j = j+1$

\EndFor {\bf end}

\EndIf {\bf end}

\State $i = i+1$

\EndFor {\bf end}

\State \Return Exhaustive set of OPF solutions ${\mathcal X}_e$

\end{algorithmic}}
\end{algorithm}




\subsection{Assessing Set Overlap}
Once a \textit{limited} set of points from the nonlinear objective optimization (Alg.~\ref{algo:iterative}) and an \textit{exhaustive} set of points from the partitioning algorithm (Alg.~\ref{algo:exhaustive}) have been obtained, a dissimilarity metric is used to compare the performance of the different functions from the Appendix. This dissimilarity metric is used to answer the following question: how well do the limited blue points span the space quantified by the small orange points in Fig.~\ref{fig:Haus}? For example, if all of the blue points are clustered into a single corner of the feasible space, then the associated nonlinear objective function performed poorly. 

\subsubsection{Hausdorff distance metric}
There are many ways to measure the ``distance" between two multi-dimensional clouds of profiles in $\mathbb{R}^n$. In this paper, we will consider the Hausdorff distance, which 
is computed as the maximum of the minimum distance (with respect to some norm) between each point in set ${\mathcal A}$ and all of the points from the other set ${\mathcal B}$~\cite{b4}:
\begin{equation}\label{eq: norm}
    H^*({\mathcal A},{\mathcal B}) = {\rm max}^{{\bm x} \in {\mathcal A}}\{{\rm min}^{{\bm y} \in {\mathcal B}}\{\left\Vert {\bm x},{\bm y}\right\Vert \}\}.
\end{equation}
Because, as previously formulated, the distance is not symmetrical, i.e., $H^*({\mathcal A}, {\mathcal B})\neq H^*({\mathcal B}, {\mathcal A})$, the distance must be computed for both directions and the maximum then taken:
\begin{equation}
    H({\mathcal A},{\mathcal B}) = {\rm max} \{H^*({\mathcal A},{\mathcal B}), H^*({\mathcal B},{\mathcal A})\}.
\end{equation}
\subsubsection{Hausdorff applied to OPF solution sets}
We use the previously computed exhaustive dataset $\mathcal{X}_e$ and compare it with datasets generated by various competing nonlinear objectives: $d_1 = H(\mathcal{S}_{f1}, \mathcal{X}_e)$, $\ldots$, $d_{40} = H(\mathcal{S}_{f40}, \mathcal{X}_e)$.
Therefore, each distance corresponds to a function that was used to obtain an associated dataset ${\mathcal S}_{f,i}$. Because the functions' primary objective is to sample the feasible space, the Hausdorff distance is the metric we use to assess how well each function performed. The distance with the lowest value will generally determine which function was most accurate in terms of exploring and representing the exhaustive set:
\begin{equation}
    d^* = {\rm min} \{ d_1, d_2, \ldots, d_{40}\}.
\end{equation}
The value $d^*$ is the minimum distance between the sampling set and the approximated feasible space across all functions.

\subsubsection{Specifying the sets ${\mathcal S}_{f,i}$ and ${\mathcal X}_{e}$} The vectors ${\bm x},{\bm y}$ in~(\ref{eq: norm}) are elements of the sets ${\mathcal S}_{f,i}$ and ${\mathcal X}_{e}$, and they can accordingly be chosen to represent power injection solutions, voltage solutions, or a mixture. Ultimately, we want to measure the ``distance" between one power flow solution and another. If we want a ``PQ" measure, then with $\bm{p}_{e}+j\bm{q}_{e}\in\mathcal{X}_{e}$ and $\bm{p}_{f}+j\bm{q}_{f} \in\mathcal{S}_{f,i}$, the norm in (\ref{eq: norm}) corresponds to
\begin{align}\label{eq: distance_edit}
\Vert \bm{p}_{e}&+j\bm{q}_{e},\bm{p}_{f}+j\bm{q}_{f}\Vert\nonumber\\  
 & =\sqrt{\sum_{k=1}^{n}\left(({p}_{e,k}-{p}_{f,k})^{2}+({q}_{e,k}-{q}_{f,k})^{2}\right)}.
\end{align}
If we want a ``PV" measure, then the reactive power is replaced with voltage magnitude. Similarly, we can just compute ``P" or ``Q" measures individually. Using this norm definition, we leverage \cite{b4} for computing the Hausdorff distance.

\section{Numerical study}\label{sec:IV}
This section uses the IEEE 3-, 5-, 14-, 30-, and 57-bus test systems from PGLib~\cite{Babaeinejadsarookolaee:2019} in order to assess how well the presented methods explore the feasible space. The solver chosen for this task is Ipopt, an open-source interior point optimizer designed to find local solutions to nonlinear problems. Ipopt is easily implemented in the Julia programming language with the use of the PowerModels library. 
Although Ipopt worked swiftly and effectively in most of the studied cases, there were some objective functions which could not be solved; these unsolved functions are 
labelled as ``DNF" in supplementary tables~\cite{github}. We note that the results in this paper are deeply related to the performance of the Ipopt solver, but we do not explore why certain functions led to optimizer error. 

Our exhaustive rejection sampling approach required significant computational power, as
the sampling strategy grows exponentially with the system size.
Moreover, we search for $N$ additional solutions within each partition that is deemed feasible, leading to a maximum of $N\cdot n^m$ total optimization problems.
Thus, this algorithm 
was implemented using DTU's High-Performance Computing (HPC) clusters~\cite{b6}. Through rejection sampling, we collected a total of 4500000 OPF solutions across all five systems.


\subsection{Simulation results}

The performance analysis of the functions consists of two parts: the first one investigates the Hausdorff distance associated with each variable individually (i.e., using just power $\bm p$ or just voltage $\bm v$ in the norm (\ref{eq: distance_edit})); the second one investigates the Hausdorff distance of the two-dimensional results of the PG-QG and PG-VM distances (i.e., using ${\bm p}+j{\bm p}$ or $[{\bm p}, {\bm v}]$ in the norm (\ref{eq: distance_edit})). These distances are considered over $N=300$ iterations of Alg.~\ref{algo:iterative}. Although all variables are expressed in p.u., the Hausdorff distance values substantially vary between different variables and systems.
Therefore, 
it is the comparison between different functions that determines the best one. 

\begin{figure}[!ht]
\centering
\includegraphics[scale=0.15]{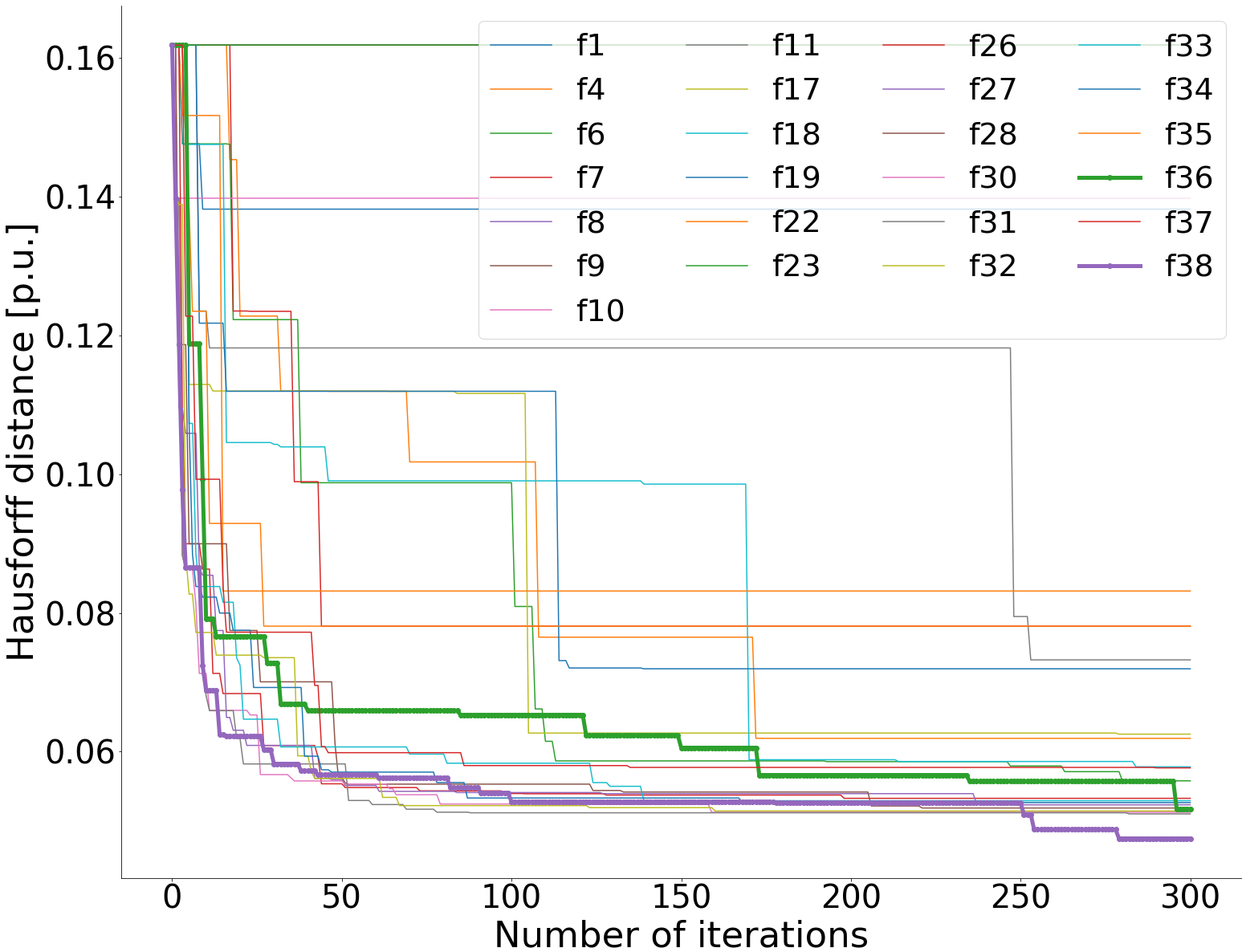}
\caption{Progression of the Hausdorff distance on the 14-bus system, as more iterations of Alg.~\ref{algo:iterative} are performed, when just voltage $\bm v$ is measured in (\ref{eq: distance_edit}). $f_{36}$ and $f_{38}$ are highlighted, and diverging functions are excluded.}
\label{hausevo14}
\end{figure}

The feasible space of voltage magnitude solutions, as depicted in Fig.~\ref{hausevo14}, is fairly well represented across all systems. For this variable, distance measures such as the Manhattan distance ($f_{30}$, $f_{31}$, $f_{32}$) and the Euclidean distance ($f_{33}$, $f_{34}$) seem to work best when a logarithmic expression is applied. Changing the base of the logarithm did not make much difference in the final values. What improved the performance of the previous voltage sampling was adding the power variables to the objectives. Hence, functions $f_{36}$ and $f_{37}$, were the best-performing ones in the list. Including power boosted the sampling to reach more remote regions in the feasible space.

The voltage angle showed the same behaviour as the voltage magnitude. The best performing functions were the ones that applied a logarithm to a distance measure that included the voltage angle. These would be functions $f_{33}$ and $f_{34}$. In this case, the benefits of adding a power component to the expressions do not seem to improve the sampling.

\begin{figure}[h!]
    \centering
    \subfloat[Function 11]{{\includegraphics[width=4.25cm]{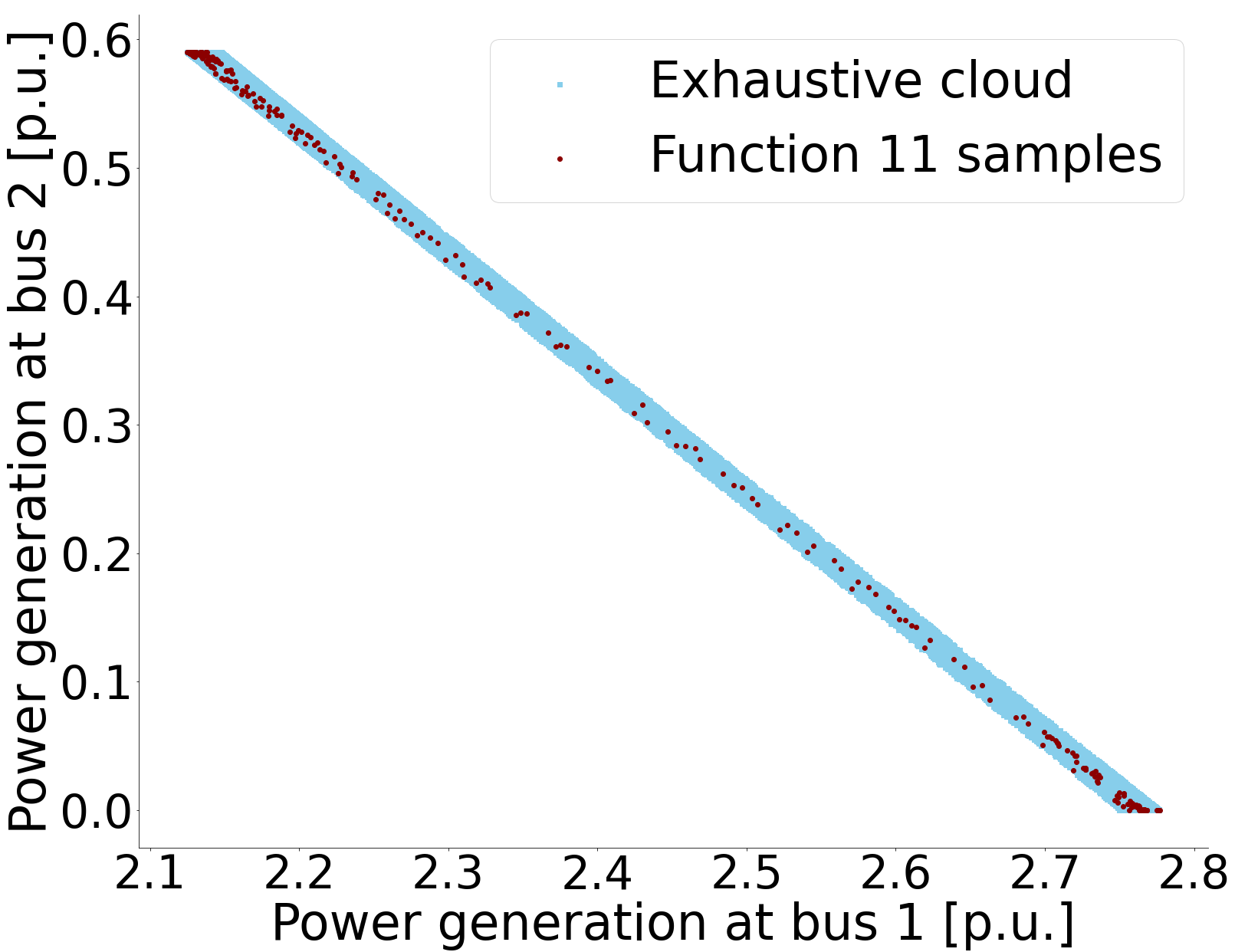}}}
    \,
    \subfloat[Function 32]{{\includegraphics[width=4.25cm]{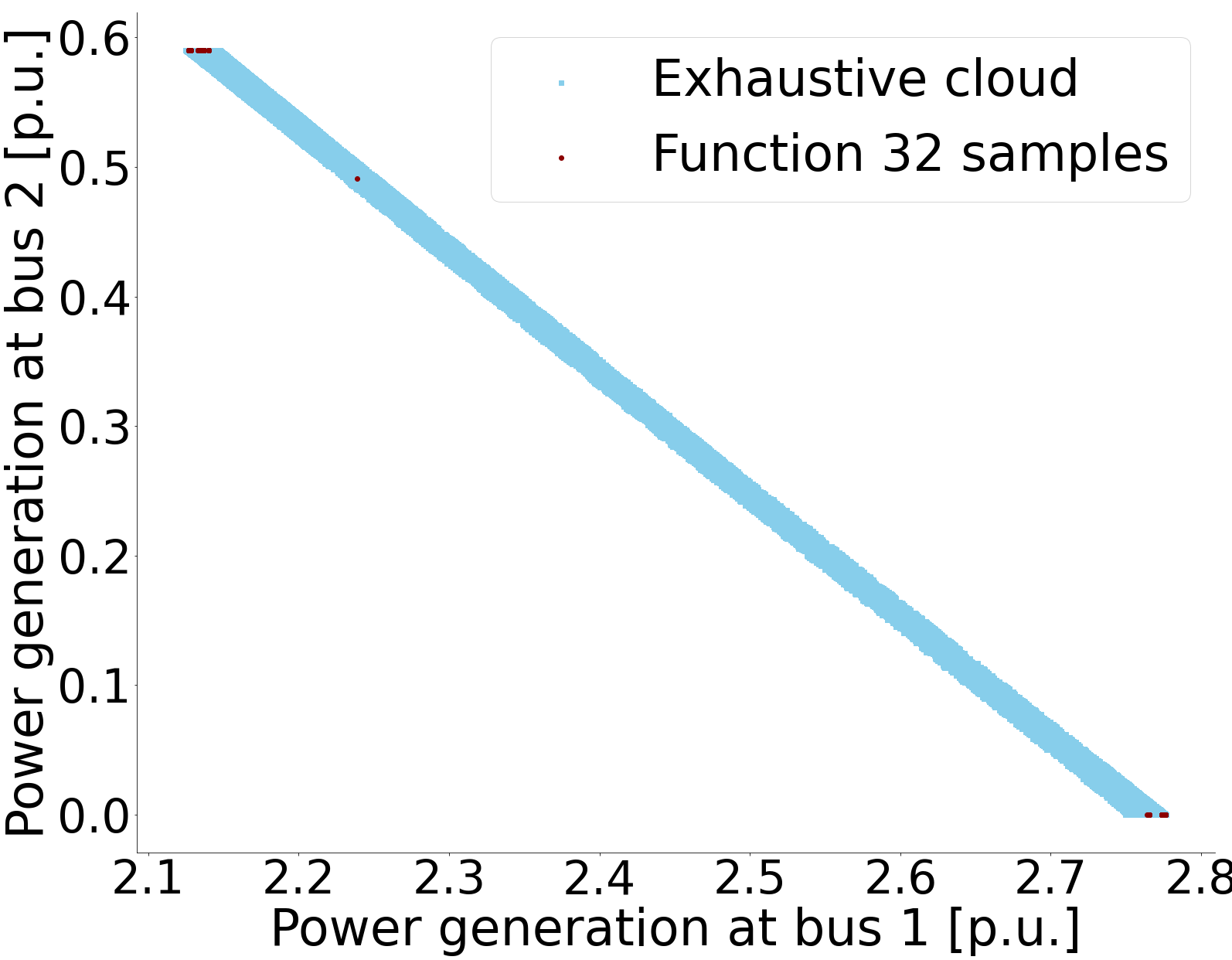} }}
    \caption{Exhaustive data is plotted as blue dots and function sampling is plotted as red dots. This figure portrays the difference between good (left) and bad (right) nonlinear objective functions by comparing the active power generation at buses 1 and 2 in the IEEE 14-bus system. In both cases, there is some clustering on the edges. However, this is much more severe for $f_{32}$.}
    \label{comparar14}
\end{figure}

\begin{figure}[h!]
\centering
\includegraphics[scale=0.15]{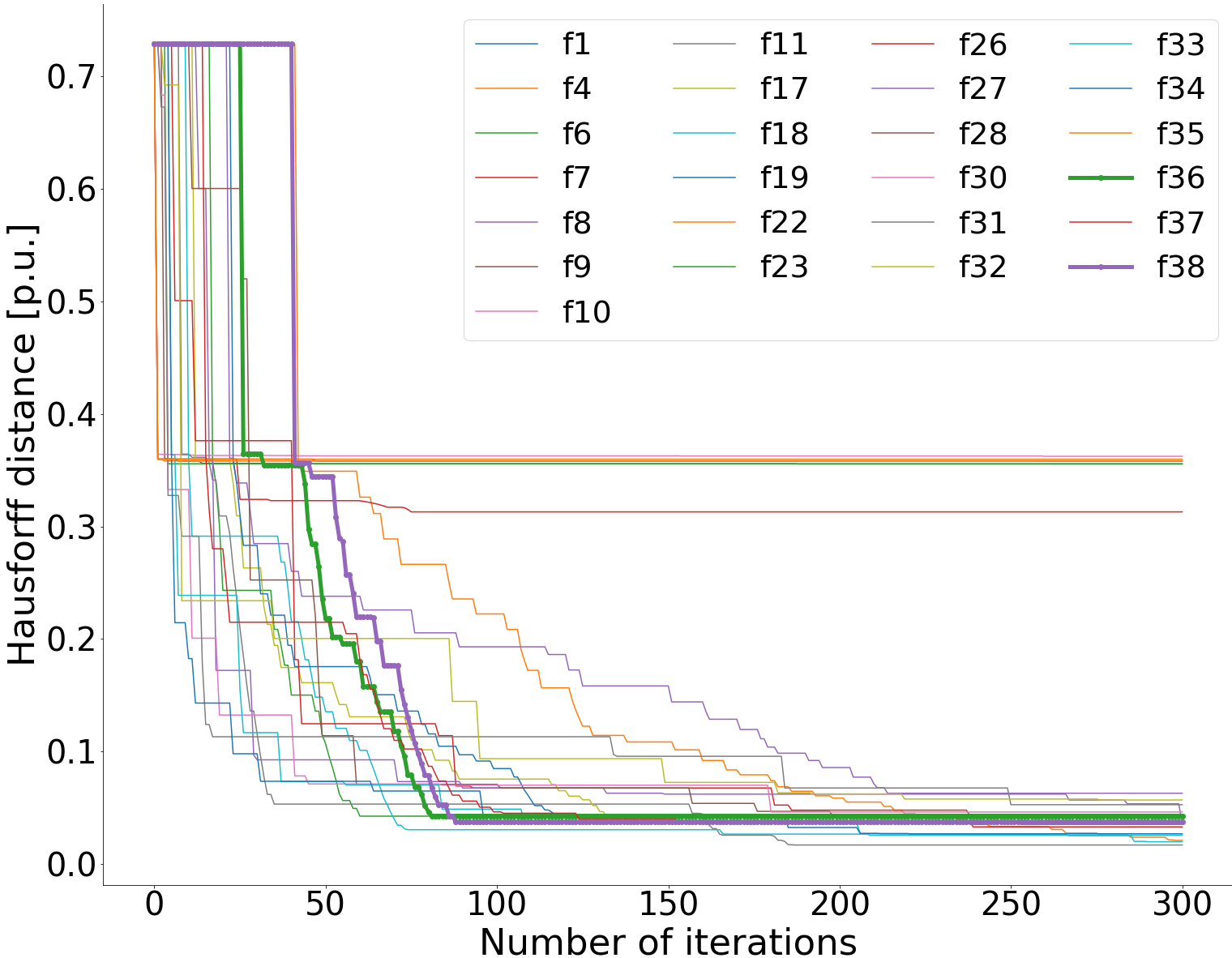}
\caption{Progression of the Hausdorff distance on the 14-bus system, as more iterations of Alg.~\ref{algo:iterative} are performed, when just power $\bm p$ is measured in (\ref{eq: distance_edit}). This figure clearly shows two different groups of functions. One group gets stuck at a value of 0.31, and a second group overcomes this first limit and reaches a lower value of 0.05. The second group samples the feasible space much more effectively. 
$f_{36}$ and $f_{38}$ are highlighted for posterior analysis.}
\label{hausevopg14}
\end{figure}

Looking at the Cartesian components of the power generation separately, no significant winner emerges. 
In this case, some groups of functions performed better than others (e.g., see Fig. \ref{comparar14}), but not strongly enough to support a conclusive statement. Thus, we extend the analysis into the complex plane. Figures \ref{hausevo14} and \ref{hausevopg14} illustrate how the Hausdorff distance of all feasible functions evolves. This metric is highly dependent on the number of iterations performed, and certain expressions get quickly stuck in some regions.

To determine the best performing functions when multiple variables are simultaneously measured (i.e., PG-QG and PG-VM distances), a point scoring system is applied. Each system awards the best ten functions with points, starting with ten to the best, and decreasing linearly until assigning one point to the tenth. Table \ref{table} expresses the final results of this scoring across all power systems. The Hausdorff distance values for \textit{all} functions can be found on our public GitHub \cite{github}.

\begin{table}[h]
\begin{center}
\caption{Five best scores for multi-variable Hausdorff distance considering complex power (left), power/voltage (middle), and the overall performance (right)}\label{table}
\begin{tabular}{c|c||c|c||c|c}
 \textbf{Func.}     &    \textbf{PQ score}    &    \textbf{Func.}    &  \textbf{PV score} & \textbf{Func}. &  \textbf{Overall score}\\
 \hline
37    & 33     & 33     & 32  & 36 & 51 \\
18    & 28     & 38     & 27  & 38 & 50 \\
36    & 25     & 34     & 26  & 37 & 48 \\
13    & 25     & 36     & 26  & 33 & 45 \\
8, 38 & 23, 23 & 12, 13 & 17  & 34 & 40
\end{tabular}
\end{center}
\end{table}

Table \ref{table} shows that $f_{36}$ and $f_{38}$ are the top two functions if we consider the points across both PQ and PV scores. These two functions apply a logarithm to the Euclidean and Manhattan distance, respectively. The difference with the other expressions that also apply a logarithm is that these two use both voltage and power, so it reaffirms the previous statement that adding multiple variables will incentivize sampling in more remote areas. Furthermore, the best performance for PQ sampling is function $f_{37}$, which applies the logarithm to the Euclidean distance of the voltage magnitude and complex power. The same function without the voltage magnitude term is the best one for PV sampling.

\section{Conclusions}\label{sec:V}
This paper investigated a set of functions which explore the feasible space of an OPF problem. We developed a method for comparing the effectiveness of these different explorations which consists of three steps:


\begin{enumerate}
    \item sequentially optimize the nonlinear objectives functions,
    \item approximate the feasible space of an OPF problem by applying an exhaustive rejection sampling strategy,
    \item assess the performance of the nonlinear objective functions using the Hausdorff distance metric.
\end{enumerate}


The primary conclusion of the studied function list is that (i) exponential functions perform very poorly, (ii) Euclidean distance tended to perform the most effectively, (iii) applying a logarithm to any distance expression is a highly effective way to explore the feasible power flow space, and (iv) the bigger the number of variables included in the objective function, the more effective the function will be in reaching remote areas of the space. The authors were surprised to see that the addition of the voltage variable $v$ to objective functions already containing $p$ and $q$ was generally quite helpful.


Specifically, we consider $f_{36}$ and $f_{38}$ to be the best performing functions; these combine the logarithm with a distance metric that includes power and voltage. They generate very effective representations of the feasible spaces of both complex power and voltage. For only power sampling, functions $f_{18}$ and $f_{13}$ are also successful. 
And if only voltage exploration is needed, then functions $f_{33}$ and $f_{34}$ are also successful. Future work will investigate the applicability of these methods to collecting data in distribution grids.

\section*{Acknowledgements}
The authors gratefully acknowledge Dr. Mohammad Rasoul Narimani, who developed the original code we leveraged for testing nonlinear objective function $f_3$ within PowerModels.

\section*{Appendix} \label{appendi}
Following is a subset of the functions experimentally tested in this paper. In the following, $\Gamma$ is the set of all previously collected data points, $\mathcal G$ is the set of dispatchable generator buses, and $\mathcal N$ is the set of all buses. Lower case variables are decision variables (e.g., $p_i$), while upper case variables are numerical data points (e.g., $P_{i,j}$). We also attempted to normalize all variables between zero and one using statistical metrics of the exhaustive set. However, this strategy is omitted since it did not offer significant improvements to the study.

\medmuskip=0mu
\thinmuskip=0mu
\thickmuskip=0mu
\allowdisplaybreaks
\footnotesize{\begin{align*}
f_{18}= & \sum_{j\in\Gamma}\Bigg(\log_{10}\sqrt{\sum_{i\in{\mathcal G}}\Big(p_{i}-P_{i,j}\Big)^{2}}+\log_{10}\sqrt{\sum_{i\in{\mathcal G}}\Big(q_{i}-Q_{i,j}\Big)^{2}}\Bigg)\\
f_{34}= & \sum_{j\in\Gamma}\Bigg(\log_{2}\bigg(\sqrt{\sum_{k\in{\mathcal N}}\Big(v_{k}-V_{k,j}\Big)^{2}}\bigg)+\log_{2}\bigg(\sqrt{\sum_{k\in{\mathcal N}}\Big(\theta_{k}-\Theta_{k,j}\Big)^{2}}\bigg)\Bigg)\\
f_{36}= & \sum_{j\in\Gamma}\Bigg(\log\bigg(\sqrt{\sum_{k\in{\mathcal N}}\Big(v_{k}-V_{k,j}\Big)^{2}}\bigg)+\log\bigg(\sqrt{\sum_{k\in{\mathcal N}}\Big(\theta_{k}-\Theta_{k,j}\Big)^{2}}\bigg)\Bigg)+\\
 & \sum_{j\in\Gamma}\Bigg(\log\bigg(\sqrt{\sum_{i\in{\mathcal G}}\Big(p_{i}-P_{i,j}\Big)^{2}}\bigg)+\log\bigg(\sqrt{\sum_{i\in{\mathcal G}}\Big(q_{i}-Q_{i,j}\Big)^{2}}\bigg)\Bigg)\\
 f_{37}=  & \sum_{j\in\Gamma}\Bigg(\log\bigg(\sqrt{\sum_{i\in{\mathcal G}}\Big(p_{i}-P_{i,j}\Big)^{2}}\bigg)+\log\bigg(\sqrt{\sum_{i\in{\mathcal G}}\Big(q_{i}-Q_{i,j}\Big)^{2}}\bigg)\Bigg)+\\
 & \sum_{j\in\Gamma}\Bigg(\log\bigg(\sqrt{\sum_{k\in{\mathcal N}}\Big(v_{k}-V_{k,j}\Big)^{2}}\bigg)\Bigg)\\
f_{38}= & \sum_{j\in\Gamma}\Bigg(\log\bigg(\sum_{i\in{\mathcal G}}\left|p_{i}-P_{i,j}\right|\bigg)+\log\bigg(\sum_{i\in{\mathcal G}}\left|q_{i}-Q_{i,j}\right|\bigg)+\\
 & \log\bigg(\sum_{k\in{\mathcal N}}\left|v_{k}-V_{k,j}\right|\bigg)\Bigg)\\
\end{align*}}

\bibliographystyle{IEEEtran}
\bibliography{SGC22_bib.bib}

\end{document}